\documentclass[sigconf]{acmart}

\renewcommand\footnotetextcopyrightpermission[1]{} 
\usepackage[utf8]{inputenc}
\usepackage{subfig}
\usepackage{bm}
\usepackage{tabularx}
\usepackage[font=small,labelfont=bf]{caption}
\usepackage{enumitem}
\usepackage{cprotect}
\usepackage{relsize}
\usepackage{threeparttable}
\usepackage{longtable}
\usepackage{booktabs}
\usepackage{multirow}
\usepackage{mathtools}
\usepackage{qtree}
\usepackage{xspace}
\usepackage{blkarray}
\usepackage{hyperref}

\newcommand{\sys}{\textsc{Wi-Fi Assist}\xspace}

\newcommand{\shortvertbreak}{\vspace{0.75 mm}}

\newcolumntype{L}{>{\centering\arraybackslash}m{2cm}}

\DeclareFixedFont{\auacc}{OT1}{phv}{m}{n}{12} 
\DeclareFixedFont{\auaccb}{OT1}{phv}{m}{n}{10} 

\makeatletter
\def\@copyrightspace{\relax}
\makeatother

\setcopyright{none}

\begin{document}

\title{\sys : Enhancing Vehicular Wi-Fi Connectivity with an Infrastructure-driven Approach}

\author{Antonio Rodrigues}
\affiliation{%
    \institution{Carnegie Mellon University}
    \city{Pittsburgh} 
    \state{PA}
    \country{USA}}
\affiliation{%
    \institution{Faculty of Engineering, Univ. of Porto}
    \country{Porto, Portugal}}
\email{adamiaonr@cmu.edu}

\author{Peter Steenkiste}
\affiliation{%
    \institution{Carnegie Mellon University}
    \city{Pittsburgh} 
    \state{PA}
    \country{USA}}
\email{prs@cs.cmu.edu}

\author{Ana Aguiar}
\affiliation{%
    \institution{Faculty of Engineering, Univ. of Porto}
    \country{Portugal}}
\affiliation{%
    \institution{Instituto de Telecomunicacoes}
    \city{Porto} 
    \country{Portugal}}
\email{anaa@fe.up.pt}

\begin{abstract}

Vehicles access the Internet via cellular networks, instead of 
Wi-Fi networks. This choice has been mostly justified by the ubiquitous 
coverage of cellular networks: Wi-Fi coverage has been shown to be 
inadequate in the past, even in urban areas. 

We argue that providing Internet connectivity to vehicles via Wi-Fi is worth 
a revisit. Motivated by improvements in Wi-Fi network coverage in recent 
years, we propose \sys, an add-on to current Wi-Fi infrastructures which 
differs from existing solutions in two key ways: (1) it is heavily infrastructure-driven; and (2) 
defines an interface for low-latency cooperation between different WLAN 
service sets, managed by different service providers.

\end{abstract}

\maketitle

\section{Introduction}
\label{sec:intro}

%
%
%
%
%
%
%

In practice, vehicles access the Internet via cellular networks, instead of 
Wi-Fi networks. Car manufacturers - e.g. BMW and Toyota - equip vehicles with cellular radios to provide 
Internet access~\cite{bmw-assist,toyota-safety-connect}. 
Wi-Fi can deliver better throughput than cellular (3G~\cite{Deshpande:2010:PCM:1879141.1879180}, 
LTE~\cite{Deng:2014:WLB:2663716.2663727}), at lower energy and monetary costs, 
and reduce the increasing strain on cellular network capacity via 
traffic offloading~\cite{6824752}.
%
%
Nonetheless, the choice falls on cellular networks due to their ubiquitous 
\textit{coverage}. Cellular base stations cover large areas ($<1$\,km to 10s\,of\,km) 
and their deployment is carefully planned by mobile carriers. This keeps 
the number of handoffs low, and allows for `make-before-break' 
handoffs while under the service of the same carrier. To a vehicle, this 
translates into continuous Internet connectivity, with few interruptions. 
In contrast, the coverage of Wi-Fi APs is narrower (10s\,of\,m), WLAN service sets 
are typically small - e.g. $<5$ APs - and their deployment is often unplanned. As 
a result, in-car Internet connectivity over Wi-Fi is highly susceptible to 
interruptions, even in urban areas. Measurement studies in urban 
areas show that vehicular Wi-Fi sessions may experience Internet connectivity 
disruptions up to 42\% of non-overlapping 1 second segments, 
with a median length of 5 consecutive segments~\cite{Deshpande:2010:PCM:1879141.1879180}. 
Other studies report median periods with no Internet connectivity up to $30$ 
seconds~\cite{Bychkovsky:2006:MSV:1161089.1161097, Eriksson:2008:CVC:1409944.1409968}. 

In this short position paper, we argue that using Wi-Fi for vehicular Internet is worth studying, 
for three reasons. First, the aforementioned measurement 
studies were conducted in the late 2000s\slash early 
2010s. Wi-Fi coverage in urban areas is likely to have improved 
significantly by now. Second, these studies only consider Internet connectivity via 
`open' Wi-Fi hotspots~\cite{Bychkovsky:2006:MSV:1161089.1161097, 
Eriksson:2008:CVC:1409944.1409968} or hotspots serviced by a 
particular provider~\cite{Deshpande:2010:PCM:1879141.1879180}. This limitation is likely 
to hide a lot of connectivity opportunities. 
Finally, most research on Wi-Fi vehicular connectivity focuses on client-centric 
approaches to reduce Wi-Fi connection setup time, overlooking the benefits of 
infrastructure-side support. Solutions such as IEEE 802.11k~\cite{4544755} and IEEE 
802.11r~\cite{4573292} do include infrastructure-side support, but 
are mostly tuned for `walking spped' scenarios~\cite{Song:2017:WGT:3098822.3098846} and 
can only be applied within a single WLAN service set. 
The infrastructure has a wider view over the whole wireless system, both temporally 
and spatially (e.g., quality of Wi-Fi signal overtime\slash at multiple 
locations). We believe that ISPs, which have some form of control over most APs 
in urban scenarios, can enhance vehicular Internet service by (1) preemptively 
configuring an appropriate set of APs in the vehicles' path; and (2) opening 
a set of cross-provider interfaces, allowing configuration state to flow 
in-between APs managed by different administrative domains. This can be 
enabled by the recently proposed concept of \textit{Network 
Service Support}~\cite{Panda:2016:ONI:2875951.2875953}, which in turn relies 
on Software Defined Networking (SDN), Network Function Virtualization (NFV) and 
Cloudlets. 


\section{Motivation}
\label{sec:motivation}




We obtain a preliminary profile of urban Wi-Fi coverage for vehicular scenarios using 
data from the SenseMyCity project~\cite{DBLP:journals/corr/RodriguesAB14}. 
This dataset contains mobility traces collected from smartphones 
of 100s of crowdsourcing volunteers in the city of Porto, Portugal. More specifically, 
we use its Wi-Fi scan dataset, which contains $\sim57$ million entries of Wi-Fi scans, 
from a total 295.000 different APs. The data includes both 
pedestrian and vehicular traces: as a conservative approach to isolate vehicular 
traces, we discard those with median speeds lower than 5.6 m\slash s ($\sim20$ km\slash h). 

\begin{figure}[t]
    \centering
        \includegraphics[width=0.475\textwidth] {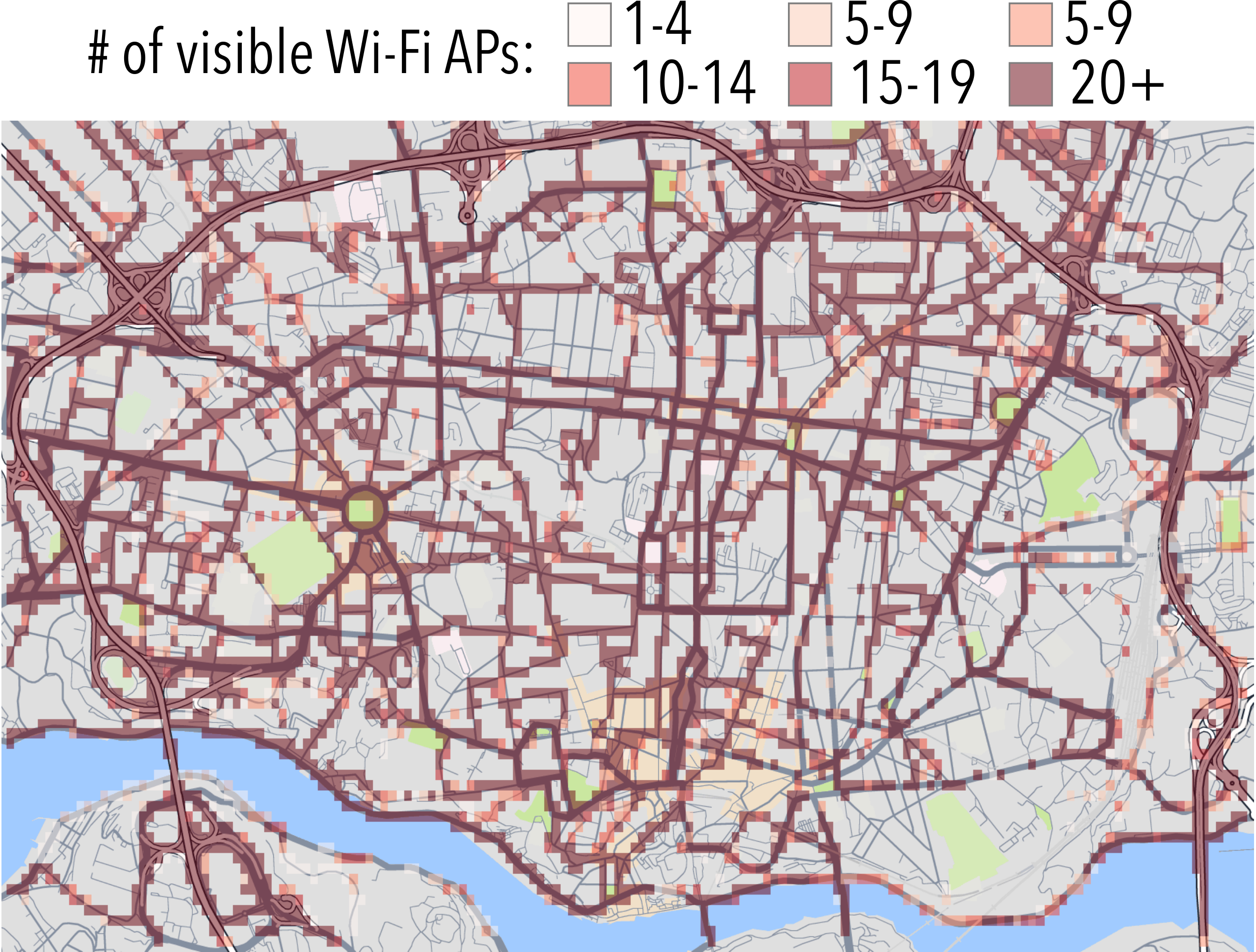}
        \cprotect\caption{Map of Porto, Portugal, divided in $50\times50$ meter cells. 
        Shading indicates number of different APs visible in the cell.}
        \label{fig:motivation-coverage}
\end{figure}


\begin{table}[t]
\centering
\caption{Data summary}
\vspace{-2mm}
\label{table:motivation-data-summary}
\newcommand{\tabitem}{~~~~~\llap{\textbullet}~}
\begin{tabular}{@{}ll@{}}
\toprule
\# of scans                                     & 95300 \\ \midrule
\# of visible APs                               & 80237 \\ \midrule
\# of ESSIDs                                    & 42843 \\
\bottomrule
\end{tabular}
\quad
\begin{tabular}{@{}lll@{}}
\toprule
\multirow{4}{*}{\# of APs per ESSID}    & 1-4   & 41988 \\
                                        & 5-9   & 684   \\
                                        & 10-49 & 150   \\
                                        & 50+   & 21    \\ \midrule
\multirow{3}{*}{\# of APs per auth.}    & Open          & 21827 \\
                                        & WPA2          & 48102 \\
                                        & Other         & 7599  \\
\bottomrule
\end{tabular}
\end{table}

Fig.~\ref{fig:motivation-coverage} shows that Wi-Fi coverage is pervasive along 
the main roads: the $50\times50$ meter cells which overlap the thicker lines 
in the roadmap report more than 20 different visible APs (more specifically, unique 
MAC addresses). Moreover, additional data in Table~\ref{table:motivation-data-summary} 
shows that most networks (identified by ESSIDs) are composed by a small number 
of APs (1 to 4), motivating the need for cooperation between networks 
managed by different WLAN service sets. There are cases of ESSIDs which aggregate a large number of 
APs - e.g. the `Fon' network with $\sim13$k different APs, or `MEO-WiFi' with 
$\sim6$k APs\footnote{Data not shown in this document.} - however, we cannot derive their degree of support 
for roaming optimizations from the data. Regarding authentication, 
the most common type is `WPA2 Personal'. This is followed by `open' APs, mostly 
part of large networks such as `Fon' or `MEO-WiFi', on which users authenticate via a login\slash password 
combo. This shows there are benefits in exploring ways of enabling APs 
protected by `WPA2-Personal' authentication, which is effectively removing 
$>\sim50$\% of connectivity opportunities for vehicles.

\section{Proposal : \sys}
\label{sec:proposal}

We propose an add-on to current Wi-Fi infrastructures - \sys - which aims at 
a general reduction of Wi-Fi connection setup time to any Wi-Fi networks found by 
moving vehicles. If coupled with 
wide adoption by ISPs, this could improve both Wi-Fi connectivity and 
throughput in vehicular scenarios, previously shown to be at odds with each 
other due to DHCP delays in Wi-Fi connection setup~\cite{Soroush:2011:CWM:2079296.2079300}.

\subsection{Overview}
\label{subsec:proposal-overview}

Despite its early stage, \sys's design differs from existing solutions in 
two key ways: (1) it is heavily infrastructure-driven; and (2) 
defines an interface for low-latency cooperation between different WLAN 
service sets, managed by different service providers.\shortvertbreak


\begin{figure}[t]
    \centering
        \includegraphics[width=0.50\textwidth] {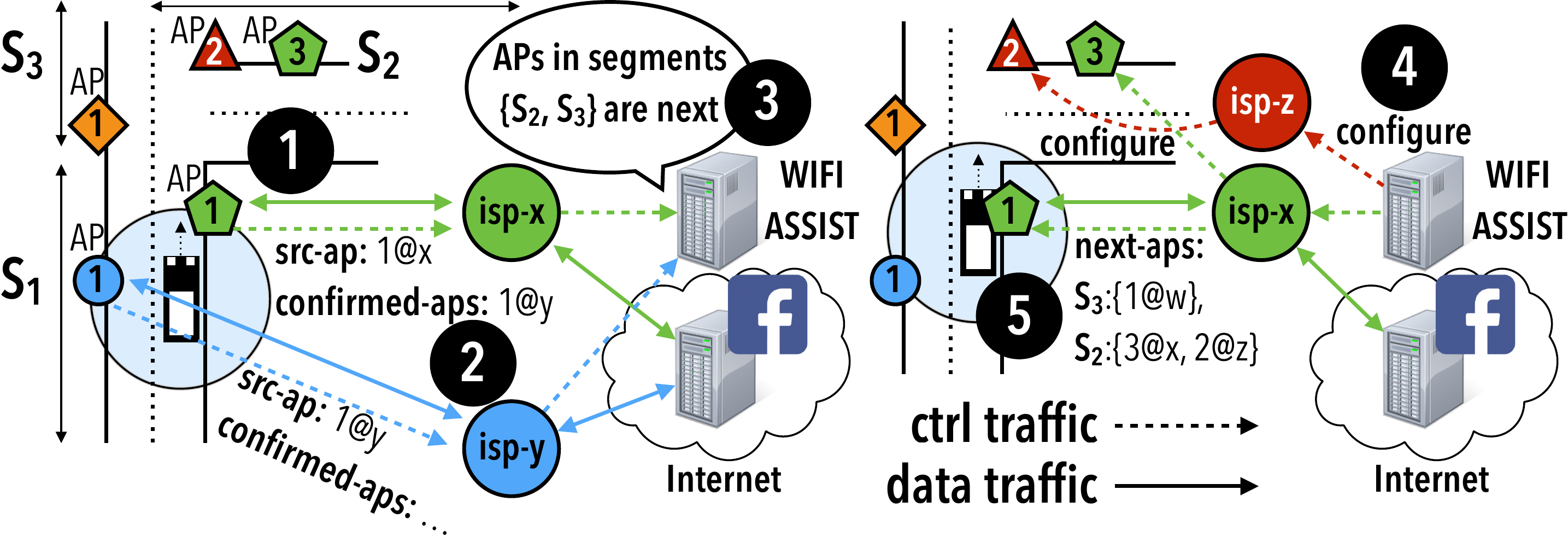}
        \cprotect\caption{Overview of \sys's design}
        \label{fig:design-proposal}
\end{figure}

\noindent\textbf{Infrastructure-driven design:} \sys aims at transferring 
work from the client to the infrastructure side, so as to shrink 
non-connectivity periods as much as possible. Non-connectivity periods have 3 
main causes: (1) lack of Wi-Fi signal coverage; (2) failure to 
connect to a Wi-Fi network; and (3) Wi-Fi connection setup. While (1) is 
a `deal-breaker', Fig~\ref{fig:motivation-coverage} shows that it may be a 
rare event. \sys mitigates (2) and (3) by preemptively 
establishing association\slash authentication state and encryption 
keys at Wi-Fi APs located along the vehicle's trajectory. Ideally, 
after a bootstrapping step, the mobile client should only deal with 
transmitting and receiving data frames, which are eventually handled by Wi-Fi 
APs within the vehicle's range. 
To accomplish this, \sys servers keep track of a vehicle's trajectory, predict the possible future ramifications 
of the current trajectory and use data aggregated from 
multiple participant APs to decide which are better suited to handle a 
particular vehicle (steps 1, 2, 3 and 5 in Fig.~\ref{fig:design-proposal}).\shortvertbreak

\noindent\textbf{Cross-provider interface:} In order to expand the space 
of APs which can handle a vehicle's communication session, \sys 
must be able to request the establishment of association\slash authentication 
state in APs belonging to different WLAN service sets, often managed by 
different providers (steps 4 and 5 in Fig.~\ref{fig:design-proposal}).

\subsection{Challenges \& Initial Design Options}
\label{subsec:proposal-challenges}


\sys's realization involves the following challenges:\shortvertbreak

\noindent\textbf{Latency minimization:} \sys adds new interfaces and system 
operations, and thus new potential sources of latency, which
must be minimized : (1) communication 
with \sys servers; (2) \sys computations; and (3) AP configuration 
operations. One option is to distribute and coordinate \sys servers across 
\textit{cloudlets}~\cite{7807196, Panda:2016:ONI:2875951.2875953} managed by the participating ISPs and positioned near 
the configurable APs. The instantiation of \sys servers could be facilitated 
through NFV. Latency of configurations internal to ISPs can be 
minimized by the pre-establishment of forwarding rules using SDN capabilities.\shortvertbreak

\noindent\textbf{Maintaining transport sessions:} As seen in Fig.~\ref{fig:design-proposal}, 
a vehicle must maintain a transport session while `hopping' over 
different APs, included in different WLAN service sets. An interesting option 
proposed by~\cite{Croitoru:2015:TWM:2789770.2789786} is the use of Multipath TCP (MPTCP)~\cite{RFC6824} 
for TCP transport, which allows for multiple sub-flows within a the same TCP session, 
between different source IP:port pairs and a single server.\shortvertbreak

\noindent\textbf{Security:} \sys poses several security challenges along 3 
axes: access control, accountability and resource protection: 

\begin{itemize}[noitemsep,topsep=0pt,leftmargin=*]
    \item \textbf{Access control:} Credentials used by vehicles to access 
        the scheduled Wi-Fi APs along their path should not be abused 
        by other entities. Furthermore, usage of ISP configuration interfaces 
        by \sys should require strict authentication policies, which must 
        be fast to minimize latency. A token-based authentication system 
        (e.g. Kerberos) is an interesting solution.
    \item \textbf{Accountability:} ISPs providing configuration interfaces 
        must be able to track\slash control traffic usage by \sys instances, for billing purposes.
    \item \textbf{Resource protection:} A malicious party - either a \sys instance, provider or vehicle - must not 
        compromise the security of data traffic nor the integrity of the network infrastructure.
\end{itemize} 


%
\bibliographystyle{myabbrv}
{\scriptsize \bibliography{conext-2017}}  
%
%

\end{document}